\documentclass[12pt,letterpaper]{article}
\usepackage{amsmath,amssymb,array,calc,rotating,epsfig,psfrag,amscd, cite}

\makeatletter
\renewcommand\section{\@startsection {section}{1}{\z@}%
                                   {-3.5ex \@plus -1ex \@minus -.2ex}
                                   {2.3ex \@plus.2ex}%
                                   {\normalfont\large\bfseries}}
\renewcommand\subsection{\@startsection{subsection}{2}{\z@}%
                                     {-3.25ex\@plus -1ex \@minus -.2ex}%
                                     {1.5ex \@plus .2ex}%
                                     {\normalfont\bfseries}}
\makeatother

\let\non\nonumber

\let\a=\alpha

\let\s=\sigma

\let\S=\Sigma

\newcommand{\bea}{\begin{eqnarray}}
\newcommand{\eea}{\end{eqnarray}}
\newcommand{\be}{\begin{equation}}
\newcommand{\ee}{\end{equation}}

\renewcommand{\O}{\operatorname{O}}

\newcommand{\m}{\mu}

\newcommand{\p}{\partial}

\newcommand{\C}[1]{$(\ref{#1})$}

\def\IZ{\relax\ifmmode\mathchoice
{\hbox{\cmss Z\kern-.4em Z}}{\hbox{\cmss Z\kern-.4em Z}}
{\lower.9pt\hbox{\cmsss Z\kern-.4em Z}} {\lower1.2pt\hbox{\cmsss
Z\kern-.4em Z}}\else{\cmss Z\kern-.4em Z}\fi}
\def\IR{\relax{\rm I\kern-.18em R}}

\def\one{{\hbox{ 1\kern-.8mm l}}}

\newlength{\bredde}
\def\slash#1{\settowidth{\bredde}{$#1$}\ifmmode\,\raisebox{.15ex}{/}
\hspace*{-\bredde} #1\else$\,\raisebox{.15ex}{/}\hspace*{-\bredde}
#1$\fi}

\newsavebox{\zzzbar}
\sbox{\zzzbar}
  {\setlength{\unitlength}{0.9em}
  \begin{picture}(0.6,0.7)
  \thinlines
  \put(0,0){\line(1,0){0.6}}
  \put(0,0.75){\line(1,0){0.575}}
  \multiput(0,0)(0.0125,0.025){30}{\rule{0.3pt}{0.3pt}}
  \multiput(0.2,0)(0.0125,0.025){30}{\rule{0.3pt}{0.3pt}}
  \put(0,0.75){\line(0,-1){0.15}}
  \put(0.015,0.75){\line(0,-1){0.1}}
  \put(0.03,0.75){\line(0,-1){0.075}}
  \put(0.045,0.75){\line(0,-1){0.05}}
  \put(0.05,0.75){\line(0,-1){0.025}}
  \put(0.6,0){\line(0,1){0.15}}
  \put(0.585,0){\line(0,1){0.1}}
  \put(0.57,0){\line(0,1){0.075}}
  \put(0.555,0){\line(0,1){0.05}}
  \put(0.55,0){\line(0,1){0.025}}
  \end{picture}}

\newcommand{\ena}{\end{eqnarray}}
\newcommand{\beqa}{\begin{eqnarray}}
\newcommand{\eeqa}{\end{eqnarray}}

\def\a{\alpha}

\def\m{\mu}

\def\s{\sigma}

\def\O{\Omega}

\def\S{\Sigma}

\renewcommand{\O}{\operatorname{O}}

\begin{document}
\begin{titlepage}

\begin{center}



\vskip 2 cm
{\Large \bf  Transcendentality violation in type IIB string amplitudes}\\
\vskip 1.25 cm { Anirban Basu\footnote{email address:
    anirbanbasu@hri.res.in} } \\
{\vskip 0.5cm  Harish--Chandra Research Institute, HBNI, Chhatnag Road, Jhusi,\\
Prayagraj 211019, India}

\end{center}

\vskip 2 cm

\begin{abstract}
\baselineskip=18pt

We analyze transcendentality for certain terms that arise in multiloop amplitudes in the low momentum expansion of the four graviton amplitude in type IIB string theory in ten dimensions, based on the constraints of supersymmetry and S--duality. This leads to several contributions that violate transcendentality beyond one loop at all orders in the low momentum expansion. We also perform a similar analysis for the five graviton amplitude, obtaining contributions that involve single--valued multiple zeta values beyond tree level.      

\end{abstract}

\end{titlepage}


\section{Introduction}

Transcendentality of various terms that appear in the correlation functions or amplitudes in perturbative quantum field theory sometimes determines the presence or absence of certain contributions, and also provides a systematic way of understanding the role zeta functions and harmonic sums play in the various expressions. While theories with large amount of supersymmetry sometimes have correlators or amplitudes which preserve  transcendentality or violate them by small amounts, this violation becomes generic as one reduces the amount of supersymmetry. It is also possible that in theories without supersymmetry, even though the correlator or amplitude contains terms that violate transcendentality, the maximally transcendental terms are closely related to their counterparts in a theory with enhanced supersymmetry. For example, maximally transcendental terms in transcendentality violating amplitudes in QCD can sometimes be related to their counterparts in $N=4$ Yang--Mills theory. Detailed multiloop calculations in field theory are crucial for this analysis.   

Similarly, it is natural to analyze transcendentality in amplitudes in perturbative superstring theory. In the absence of data involving multiloop string amplitudes where not much is known, it is difficult to perform a direct analysis as in field theory. In this paper, we shall look at simple patterns of transcendentality and its violation in type IIB string theory in ten dimensions which possesses maximal supersymmetry. Our analysis is primarily based on the constraints imposed by supersymmetry and S--duality of the theory, and avoids explicit calculation of multiloop amplitudes. We shall show that based on data involving only the BPS interactions in the effective action, supersymmetry and S--duality predicts transcendentality violation beyond one loop for a class of non--BPS interactions at all orders in the $\alpha'$ expansion, even though for none of these non--BPS interactions do  we have a detailed understanding of their perturbative properties from the worldsheet. We shall see how this spacetime analysis is linked to the worldsheet analysis, in particular, this leads to a definite pattern of contributions that arise from the one loop string amplitude. 

We first discuss this issue for the case of the four graviton amplitude in the type IIB theory, which leads to terms of the form $D^{2k} \mathcal{R}^4$ in the effective action, where $D$ schematically represents a derivative, while $\mathcal{R}$ represents the Riemann tensor. We also consider non--analytic terms of the schematic form $({\rm ln}D^2)^n D^{2k} \mathcal{R}^4$ in the effective action where $n$ is a positive integer, and analyze patterns of transcendentality and its violation.  
We then briefly consider the five graviton amplitude, which leads to similar conclusions. For interactions of the form $D^{2k} \mathcal{R}^5$ in the effective action that follow from the five graviton amplitude, we use supersymmetry and S--duality to predict the presence of single--valued multiple zeta values beyond tree level in the various interactions.  

In our analysis, we always drop various numerical factors of vanishing transcendentality, and keep the ones that have non--vanishing transcendentality. These include the Riemann zeta function $\zeta(a)$ which has transcendentality or weight $a$. In fact, we only need to consider the cases where $a$ is a positive integer greater than 1\footnote{Thus from $\zeta(2)=\pi^2/6$, we see that $\pi$ has weight 1.}. We also consider the multiple zeta value or the Euler--Zagier sum
\be \zeta(a_1,\ldots, a_l) = \sum_{n_1 > \ldots > n_l \geq 1; n_i \in \mathbb{Z}} \frac{1}{n_1^{a_1}\ldots n_l^{a_l}},\ee
which has weight $w = \sum_{i=1}^l a_i$ and depth $l$. In our analysis, we only need to consider cases where $a_i$ are positive integers with $a_1 >1$. Finally, we also consider the single--valued multiple zeta value $\zeta_{sv}(a_1,\ldots, a_l)$ which has weight $w = \sum_{i=1}^l a_i$ and depth $l$, and which can be expressed in terms of multiple zeta values.   

As stated above, in the absence of sufficient data about perturbative superstring amplitudes, we shall analyze transcendentality violation using supersymmetry and S--duality. For this purpose, it is useful to analyze the effective action of the type IIB theory in ten dimensions. This can be expanded in powers of $\alpha'$, and takes the form
\be \label{act}\alpha'^4 S = S^{(0)} +\sum_{k=0}^\infty S^{(3+k)}\ee
in the Einstein frame, where $S^{(0)}$ is the supergravity action given by
\be \label{sugra}S^{(0)} \sim \int d^{10}x \sqrt{-g}R  +\ldots.\ee
Considering only the local terms, $S^{(3+k)}$ includes terms of the form\footnote{In fact, the ten derivative terms in the action given by $S^{(4)}$ all vanish on--shell.}  
\be \label{local}S^{(3+k)} \sim \alpha'^{3+k}\int d^{10}x \sqrt{-g}\Big[ F_{3+k} (\Omega,\overline\Omega)D^{2k} \mathcal{R}^4+ {\widetilde{F}}_{3+k} (\Omega,\overline\Omega)D^{2(k-1)} \mathcal{R}^5+\ldots\Big],\ee
where $F_{3+k} (\Omega,\overline\Omega), {\widetilde{F}}_{3+k} (\Omega,\overline\Omega)$ are S--duality (or $SL(2,\mathbb{Z})$) invariant moduli dependent couplings, where $\Omega = \Omega_1 +i\Omega_2 = C_0 + i e^{-\phi}$. Among the moduli, $C_0$ is the R--R pseudoscalar and $\phi$ is the dilaton. On performing a weak coupling expansion around $\Omega_2 \rightarrow \infty$, $F_{3+k}(\Omega,\overline\Omega)$ and ${\widetilde{F}}_{3+k}(\Omega,\overline\Omega)$ yield several terms that are power behaved in $\Omega_2$ as well as terms involving ${\rm ln}\Omega_2$. There are also contributions that are exponentially suppressed in $\Omega_2$ which we shall ignore throughout our analysis which involve D(anti)--instanton contributions, as we are interested in analyzing transcendentality in the perturbative part of the amplitude.     

Apart from the terms in \C{local}, the effective action \C{act} also contains non--local terms.  These interactions have extra factors schematically of the form $({\rm ln}D^2)^n$ where $n$ is a positive integer. Their role in the analysis of transcendentality is determined by the local interactions which have moduli dependence of the form ${\rm ln}\Omega_2$ which we shall discuss separately.

Note that on converting to the string frame, the effective action \C{sugra} yields
\be S^{(0)} \sim \int d^{10}x \Omega_2^2 \sqrt{-g_\s}R_\s  +\ldots,\ee
while the local terms  \C{local} yield
\be \label{stringframe}S^{(3+k)} \sim \alpha'^{3+k}\int d^{10}x \sqrt{-g_\s}\Omega_2^{(1-k)/2}\Big[ F_{3+k} (\Omega,\overline\Omega)D_\s^{2k} \mathcal{R}_\s^4+ {\widetilde{F}}_{3+k} (\Omega,\overline\Omega)D_\s^{2(k-1)} \mathcal{R}_\s^5+\ldots\Big],\ee
where $\s$ denotes quantities evaluated using the string frame metric. Similarly, on converting the non--local terms to the string frame and adding the local and non--local contributions together, in the string frame all terms involving factors of ${\rm ln}\Omega_2$ cancel. In fact every term that is power behaved in $\Omega_2$, takes the form $\Omega_2^{2(1-g)}$ that can be interpreted as a perturbative $g$ loop string amplitude. 

\section{Transcendentality and its violation for the four graviton amplitude}

To begin with, we analyze the structure of transcendentality and its violation for the four graviton amplitude in type IIB string theory in ten dimensions. We first briefly list the perturbative data that is known for this amplitude at various loops from direct worldsheet calculations. 

\subsection{Perturbative data and transcendentality violation}

The tree level amplitude is given by
\be \label{tree}\mathcal{A}_4^{(0)} \sim \frac{e^{-2\phi}}{\a'^3 stu} e^{\sum_{n=1}^\infty \zeta(2n+1)\a'^{2n+1}\s_{2n+1}}\mathcal{R}^4,\ee
where
\be \s_n = s^n + t^n + u^n.\ee 
Here $s,t$ and $u$ are the Mandelstam variables given by $s = -(k_1 + k_2)^2/4, t = -(k_1+k_4)^2/4$ and $u=-(k_1 + k_3)^2/4$.
The momenta $k_i$ satisfy the constraints $\sum_{i=1}^4 k_i^\m =0$ and $k_i^2=0$. 
Assigning transcendentality or weight $-1$ to $\a' \p^2$~\footnote{This is equivalent to assigning weight $-1$ to the dimensionless Mandelstam variable as in~\cite{DHoker:2019blr}. } and $-1$ to $e^{-2\phi}$, we see that $\mathcal{A}_4^{(0)}$ has weight 2 and every term in the $\alpha'$ expansion preserves transcendentality.   

The one loop amplitude can be expressed as the sum of analytic and non--analytic contributions in the external momenta, and thus~\cite{Green:1999pv,Green:2008uj,DHoker:2015gmr,DHoker:2019blr} 
\be \mathcal{A}_4^{(1)} =\mathcal{A}_{an}^{(1)} + \mathcal{A}_{non-an}^{(1)} .\ee 
The analytic contribution (which yields local terms in the effective action) is known upto the $D^{12}\mathcal{R}^4$ term in the low momentum expansion and is given by
\be \mathcal{A}_{an}^{(1)} \sim\zeta(2)\Big[1 + \zeta(3) \a'^3\s_3+ \zeta(5)\a'^5\s_5 + \zeta(3)^2 \a'^6 \s_3^2 \Big] \mathcal{R}^4.\ee
The non--analytic contribution (which yields local and non--local terms in the effective action in the Einstein frame) can be evaluated at all orders~\cite{DHoker:2019blr}, and the leading terms in the $\alpha'$ expansion are schematically given by\footnote{Also see ~\cite{Green:2006gt,Green:2008bf,Basu:2008cf,Basu:2013goa,Alday:2018pdi} for relevant discussions.}
\be \label{nonan}\mathcal{A}_{non-an}^{(1)} \sim\zeta(2)\Big[ \a' s {\rm ln} (\a' \m_1^{-2} s) + \zeta(3) \a'^4 s^4 {\rm ln}(\a' \m_4^{-2} s) + \zeta(5) \a'^6 s^6 {\rm ln}(\a'\m_6^{-2} s) \Big]\mathcal{R}^4,\ee
where $\m_l$ is a dimensionless renormalization scale. Assigning weight 1 to ${\rm ln}(\a'\m_l^{-2} \p^2)$~\cite{DHoker:2019blr} we see that every term in $\mathcal{A}_{an}^{(1)}$ and $\mathcal{A}_{non-an}^{(1)}$ has weight 2. Based on these observations, it is natural to expect that transcendentality is preserved for the one loop amplitude. 
Assuming this, we see that transcendentality is preserved upto one loop for the four graviton amplitude.  

Now let us consider the analytic terms that are known in the low momentum expansion of the two loop amplitude. They are given by~\cite{DHoker:2005jhf,DHoker:2014oxd}
\be \mathcal{A}^{(2)}_{4}\sim e^{2\phi}\zeta(4)\Big[ \a'^2\s_2 +  \a'^3\s_3\Big]\mathcal{R}^4.\ee 
The two terms have weights 3 and 2 respectively, and thus we see that transcendentality is violated at two loops. Similarly, at three loops the $D^6\mathcal{R}^4$ term is given by~\cite{Gomez:2013sla}
\be  \mathcal{A}^{(3)}_{4}\sim e^{4\phi}\zeta(6)\a'^3 \s_3\mathcal{R}^4\ee
which has weight 5, hence violating transcendentality.  

Thus we see that transcendentality is violated beyond one loop for the four graviton amplitude. This leads us naturally to two questions. First, can we analyze the transcendental structure at one loop beyond the first few orders in the low momentum expansion, and see if there is any violation? Secondly, can we understand the violation at two loops and beyond at higher orders in the low momentum expansion? We now turn to this analysis based on spacetime, rather than worldsheet, techniques.     

Note that for the local contributions we have assigned weight $-1$ to $\a' \p^2$. Thus all the spacetime interactions that arise at the same order in the $\alpha'$ expansion have the same transcendentality (for example, $s^3\mathcal{R}^4$, $\tilde{s}^2\mathcal{R}^5$ and $\mathcal{F}_5^2 \mathcal{R}^6$ where $\tilde{s}$ is a Mandelstam variable for the five graviton amplitude, and $\mathcal{F}_5$ is the self--dual field strength)\footnote{For the purposes of determining weights, we are counting powers of $\a'$ with respect to $\mathcal{R}^4$ which is clear from the amplitudes above.}. The moduli dependent couplings of these interactions have non--vanishing weight on assigning weight $-1/2$ to $\Omega_2$. Thus it is natural to assign weight $-1/2$ to $\Omega_1$ as well, and vanishing weight to all other fields. Hence among the fields, apart from the overall moduli dependent couplings, the fields in the spacetime interactions have no weight. This is also true if the interactions depend on $\Omega$ because such terms can only involve factors of $\Omega_2^{-1} \p_\m\Omega$.  

On the other hand, our analysis based on supersymmetry and S--duality is in the Einstein frame where for example, the metric has non--vanishing weight, and also one gets factors of ${\rm ln}\Omega_2$. Hence we analyze transcendentality only in the string frame, where couplings admit a perturbative expansion of the form $\Omega_2^{2(1-g)}$.    

Thus we have assigned non--vanishing weight only to $\a' \p^2$ (and also to ${\rm ln}(\a' \p^2)$) and $\Omega$, which are related to the worldsheet and spacetime complexified couplings respectively, which is natural from the point of view of field theory where one can similarly analyze the issues of transcendentality and its violation. For example, assigning weight $-1$ to the Yang-Mills coupling in $N=4$ Yang--Mills, we see that every term in the perturbative expansion of the planar cusp anomalous dimension has weight $-2$~\cite{Beisert:2006ez,Kotikov:2006ts}. On the other hand for example, transcendentality is violated in the perturbative expansion of the planar anomalous dimension of the Konishi operator (see~\cite{Leurent:2013mr} for the eight loop result).

\subsection{Supersymmetry and S--duality constraints on transcendentality violation}

From the previous analysis, we see that is not straightforward to obtain perturbative data beyond the first few loops from worldsheet calculations. Even for those cases where the amplitude is known, it is non--trivial to obtain data at arbitrary orders in the low momentum expansion. So in this section we shall proceed differently.  

In worldsheet calculations, at every loop order we obtain an expression for all $\alpha'$. Expanding in small $\alpha'$, this leads us to the higher derivative interactions in the effective action. A complementary point of view is to analyze the effective action, where at a fixed order in the $\alpha'$ expansion, every interaction has a moduli dependent coupling that is exact in the string coupling, even nonperturbatively. In the Einstein frame, for the purely gravitational terms in the effective action, these moduli dependent couplings are S--duality (or $SL(2,\mathbb{Z})$) invariant. We shall see that constraints of supersymmetry and S--duality impose restrictions on these couplings, which when expanded for weak string coupling, give us information about the transcendental nature of the perturbative data.      

The action \C{act} is invariant under the supersymmetry transformation
\be \delta  = \delta^{(0)} +\sum_{k=0}^\infty \delta^{(3+k)}\ee
and hence $\delta S=0$. Here $\delta^{(0)}$ is the supersymmetry transformation of supergravity, while the other terms $\delta^{(3+k)} \sim \alpha'^{3+k}$ yield corrected supersymmetry transformations\footnote{Note that $\delta^{(4)}$ vanishes on--shell, since $S^{(4)}$ vanishes on--shell.}. Expanding  $\delta S=0$ in powers of $\alpha'$, while the leading contribution gives us $\delta^{(0)} S^{(0)}=0$ , the subleading contributions give us that
\be \label{noether}\delta^{(0)} S^{(n)}  +\delta^{(n)} S^{(0)} +\sum_{p+q=n}\delta^{(p)} S^{(q)}=0\ee
where $n \geq 3$, and $p, q \geq 3$. For the $1/2$ BPS $\mathcal{R}^4$ and the $1/4$ BPS $D^4\mathcal{R}^4$ interactions, only the first two terms in \C{noether} contribute corresponding to $n=3$ and $n=5$ respectively, leading to couplings that satisfy Laplace equation on moduli space. In fact, the analysis is simpler on considering the maximally fermionic interactions in the $1/2$ and $1/4$ BPS supermuliplets~\cite{Green:1998by,Sinha:2002zr}, which lead to $SL(2,\mathbb{Z})$ covariant couplings, while the other interactions are related by supersymmetry. On converting to the string frame, for the $R^4$ interaction, we obtain the perturbative contributions having weight 2~\cite{Green:1997tv} 
\be \Big[e^{-2\phi} \zeta(3) +\zeta(2)\Big]\mathcal{R}^4.\ee      
On the other hand, for the $D^4\mathcal{R}^4$ interaction, we obtain the perturbative contributions~\cite{Green:1999pu}
\be \a'^2 \Big[ e^{-2\phi} \zeta(5) + e^{2\phi} \zeta(4)\Big]\s_2\mathcal{R}^4,\ee
leading to transcendentality violation at two loops by 1 unit. 

For the $1/8$ BPS $D^6\mathcal{R}^4$ interaction as well as for the non--BPS interactions that occur at higher orders in the $\alpha'$ expansion, the last term in \C{noether} contributes. This leads to Poisson equations on moduli space, where the source terms are given by couplings at lower orders in the $\alpha'$ expansion. These source terms that arise from the supervariation of the form $\delta^{(p)}S^{(q)}$ are at least quadratic in the couplings, where one factor arises from the coupling in $S^{(q)}$ while the other factor arises from the coupling in the corrected supersymmetry transformation $\delta^{(p)}$ whose form is dictated by the closure of the supersymmetry algebra~\cite{Basu:2008cf}\footnote{For the $D^6\mathcal{R}^4$ interaction in $S^{(6)}$, the source term is given by the square of the $\mathcal{R}^4$ coupling, which is in $S^{(3)}$.} and involves an appropriate coupling in $S^{(p)}$. For the $D^6\mathcal{R}^4$ interaction, this leads to the perturbative contributions~\cite{Green:2005ba}\footnote{The structure of such Poisson equations has been deduced in~\cite{Basu:2008cf} for the maximally fermionic interactions in the various supermultiplets, and the structure of the equations for the $D^{2k}\mathcal{R}^4$ interactions should follow from supersymmetry, though this has not been elucidated in detail.}
\be \a'^3\Big[ e^{-2\phi} \zeta(3)^2 +\zeta(2)\zeta(3) +e^{2\phi} \zeta(4) +e^{4\phi}\zeta(6)\Big]\s_3\mathcal{R}^4,\ee      
leading to transcendentality violation at three loops by 3 units. Thus for these BPS interactions we obtain agreement with the structure obtained by worldsheet techniques. 

\subsubsection{Transcendentality of non--BPS interactions from recursion relations}

The $D^{2k} \mathcal{R}^4$ interactions in the effective action for $k \geq 4$ are non--BPS, and for them exact Poisson equations satisfied by the couplings are not known. They are not expected to satisfy perturbative non--renormalization theorems, and hence they are not expected to satisfy simple Poisson equations. Moreover, the coupling can split into a sum of terms each of which satisfies a separate Poisson equation. Thus for such interactions, we shall not be able to analyze all the perturbative contributions, and hence transcendentality in complete detail. However, we now show that one can still partially analyze the transcendental structure of various multiloop amplitudes that arise from these interactions at arbitrary orders in the $\alpha'$ expansion, based on our knowledge only of the BPS interactions we have discussed previously. Thus this analysis based on supersymmetry and S---duality is distinct from the worldsheet perspective.  

A Poisson equation satisfied by these couplings is of the general form
\be \label{main}4\Omega_2^2\frac{\p^2f_i}{\p\Omega \p \overline\Omega} = s_i(s_i-1) f_i + g_i,\ee
where the coupling of an interaction is given by $\sum_i f_i (\Omega,\overline\Omega)$, and $g_i (\Omega,\overline\Omega)$ is the source term in the equation for $f_i$.  We do not know into how many $f_i$ each spacetime coupling decomposes into, and for each $i$, we do not have a detailed understanding of the source term $g_i$. In fact, along with other contributions, we expect that  the source terms in the Poisson equations for the $D^{2k}\mathcal{R}^4$ interaction will also involve contributions from the $D^{2k'}\mathcal{R}^p$ couplings for $p > 4$ as well. Such kind of operator mixing is much like what happens in quantum field theory\footnote{This does not happen for the BPS interactions as all interactions that preserve the same amount of supersymmetry are expected to lie in the same short supermultiplet. Thus for example, the $D^6\mathcal{R}^4$, $D^4 \mathcal{R}^5$ and $\mathcal{R}^7$ couplings are expected to be proportional to each other and lie in a single $1/8$ BPS supermultiplet. However for non--BPS interactions, we do expect multiplet splitting.}.        

However, we can still analyze the issue of transcendentality in considerable detail for the perturbative part of $D^{2k}\mathcal{R}^4$ interaction by keeping only those source terms in the Poisson equations that arise from the $D^{2k'}\mathcal{R}^4$ interactions for $k' < k$. At sufficiently low orders in the $\alpha'$ expansion, these source terms are given by BPS interactions for which we know the complete perturbative behavior. 

Though this ignores various other source terms of the type described above, this will lead to results for the transcendentality for various multiloop amplitudes for the $D^{2k}\mathcal{R}^4$ interaction, which are difficult to derive otherwise. But now we can recursively use these results to analyze interactions at higher orders in the $\alpha'$ expansion given by $D^{2k'}\mathcal{R}^4$ for $k' > k$, where the source term involves the $D^{2k}\mathcal{R}^4$  coupling. This allows us to calculate the transcendentality of various amplitudes at arbitrary orders in the $\alpha'$ expansion.           

To understand the perturbative part of the coupling, we focus on only the $\Omega_1$ independent part of $g_i$ that is power behaved as well as possibly logarithmic in $\Omega_2$.  We ignore terms that are exponentially suppressed in $\Omega_2$ since they correspond to D--instanton anti--instanton bound states carrying vanishing R--R charge. Thus keeping only these contributions, \C{main} reduces to
\be \label{poisson}\Omega_2^2\frac{\p^2f_i}{\p\Omega_2^2} = s_i(s_i-1) f_i + \sum_m a_{m;i} \Omega_2^m +\sum_{n,p} b_{n,p;i} \Omega_2^n ({\rm ln}\Omega_2)^p.  \ee
In the source terms, $m$ and $n$ can be fractions, while $p$ is a positive integer. While the origin of terms that are purely polynomials in $\Omega_2$ is obvious, we briefly review the origin of terms that involve ${\rm ln}\Omega_2$ in the Poisson equations.     

Worldsheet calculation of loop amplitudes produce non--analytic terms in the external momenta, some examples of which have been mentioned previously at one loop in \C{nonan}. This schematically leads to interactions involving factors of $[{\rm ln}(\a'\m^{-2} s)]^n$ (where $n$ is a positive integer) in the amplitude. On converting to the Einstein frame, this leads to factors of $[{\rm ln}(\a'\m'^{-2}\hat{s}) + ({\rm ln}(\Omega_2\m^{-4} \m'^{4}))/2]^n$ for those interactions in the effective action, where $\hat{s}$ is a Mandelstam variable in the Einstein frame. Thus apart from non--analytic terms in the effective action, this also leads to an analytic term involving $[({\rm ln}(\Omega_2\m^{-4} \m'^{4}))/2]^n$. Thus source terms that are logarithmic in $\Omega_2$ in the Poisson equations arise from non--analytic terms in the string frame, and play a crucial role in S--duality. 

Now consider the equation \C{poisson}. We can write the solution in the form
\be \label{soln}f_i \sim c_{1i}\Omega_2^{s_i} +c_{2i}\Omega_2^{1-s_i} + \sum_m \widetilde{a}_{m;i} \Omega_2^m +\sum_{n,q} \widetilde{b}_{n,q;i} \Omega_2^n ({\rm ln}\Omega_2)^q .\ee
The first two terms are the solutions to the homogeneous equation, while the others yield the particular solution. The third term involving the summation depends only on the coefficients $a_{m;i}$ and not on $b_{n,p;i}$, in fact, $\widetilde{a}_{m;i}$ is proportional to $a_{m;i}$ upto numerical factors of vanishing transcendentality. These terms that are purely power behaved in $\Omega_2$ lead to local terms in the amplitude on converting to the string frame. 

What about the structure of the amplitude that results from the remaining term in \C{soln}? This term involving the summation depends only on the coefficients $b_{n,p;i}$ and not on $a_{m;i}$. In \C{poisson}, for a fixed $n$, let $p$ range from 1 to $p_{max}$. Then in \C{soln} for that value of $n$, the integer $q$ ranges from 0 to $p_{max}$. Thus while $\widetilde{b}_{n,p_{max};i}$ is proportional to $b_{n,p_{max};i}$ upto numerical factors of vanishing transcendentality, $\widetilde{b}_{n,q;i}$ for $ 0 \leq q \leq p_{max}-1$ is a linear combination of $b_{n,p;i}$ for $1 \leq p \leq p_{max}$, with coefficients of vanishing transcendentality as well\footnote{These conclusions follow from the integral
\be \int dx x^n ({\rm ln} x)^m = \frac{x^{n+1}}{m+1}\sum_{k=0}^m (-1)^k(m+1)m(m-1)\ldots(m-k+1) \frac{({\rm ln}x)^{m-k}}{(n+1)^{k+1}}\ee
which arises in calculating the particular solution of \C{poisson}.}. This statement is true for every $n$. For a fixed $n$ these terms can lead to local (from $q=0$) as well as non--local terms in the string amplitude, where the non--local terms have momentum dependence of the schematic form $({\rm ln}(\a's))^k$ for $1 \leq k \leq p_{max}$ while the loop dependence is determined by $\Omega_2^n$ on converting to the string frame. Now it is possible that the terms involving lower powers of $k$ can be absorbed into the terms involving higher powers of $k$ by changing the renormalization scale and subtracting a constant term which changes the coefficient of the local term\footnote{For example, $({\rm ln}(\a's))^2 + 2 ({\rm ln}(\a's)) = [{\rm ln}(e\a' s)]^2-1$.}. Whether this happens or not can be determined only by knowing the details of the various coefficients, which we do not have access to. However, it follows from our argument that the source terms do lead to non--local terms of the form $({\rm ln}(\a's))^{p_{max}}$ for every $n$ with its coefficient given by $b_{n,p_{max};i}$ upto numerical factors of vanishing transcendentality{\footnote{Note that if $m= s_i$ or $1-s_i$ in \C{poisson} then we also get a term of the form $\Omega_2^m {\rm ln}\Omega_2$ in the solution apart from those in \C{soln}. Similarly, if $n=s_i$ or $n=1-s_i$ in \C{poisson}, we also get an additional term of the form $\Omega_2^n ({\rm ln}\Omega_2)^{p_{max}+1}$ in \C{soln}. These degenerate cases have to be considered separately, and we do not consider them further.}. 

Thus from the structure of the source terms that contribute to the Poisson equation, we see that

(i) every source term of the form $a_n \Omega_2^n$ yields a term in the solution to the Poisson equation of the form $a_n \Omega_2^n$ on ignoring numerical factors of vanishing transcendentality, which leads to analytic terms in the amplitude,

(ii) every source term of the form $\Omega_2^n \sum_{p=1}^{p_{max}} b_{n,p}({\rm ln}\Omega_2)^p$ yields a term in the solution of the form $\Omega_2^n b_{n,p_{max}} ({\rm ln} \Omega_2)^{p_{max}}$  on ignoring numerical factors of vanishing transcendentality, which leads to a non--analytic term of the schematic form $b_{n,p_{max}} ({\rm ln}(\a's))^{p_{max}}$ in the amplitude, where the number of loops is determined by $\Omega_2^n$ on converting to the string frame,

(iii) these conclusions are independent of the eigenvalue.  

Thus even with a limited knowledge of source terms in the Poisson equations, we can obtain partial results at various loops for string amplitudes, and can we can analyze transcendentality violation for them. All other source terms will give additional contributions beyond the ones we consider.  

\subsubsection{Transcendentality and its violation: examples and generalities}

From the above discussion, we see how the source terms yield contributions at various loops to the string amplitude from the structure of the Poisson equation. In fact since we are interested in only the transcendentality, from (i), (ii) and (iii) above, it is easy to write down these contributions once we know the large $\Omega_2$ expansion of the source terms, as the solution to the eigenvalue equation essentially yields the source terms themselves, with an elementary restriction on the logarithmic dependence. For a set of source terms that arise in the Poisson equation for $F$, we label the solution including these contributions as $\left[F\right]$~\footnote{For example, if the source terms in the Poisson equation for $F$ are of the form $\zeta(6)\Omega_2^4 + \zeta(3)^2\Omega_2^2 [({\rm ln}\Omega_2)^3 +{\rm ln}\Omega_2]$, then
\be \left[F\right] \sim \zeta(6)\Omega_2^4 + \zeta(3)^2 \Omega_2^2 ({\rm ln}\Omega_2)^3.\ee}.    
We note that for the $D^{2k}\mathcal{R}^4$ couplings in \C{local} for $k \leq 3$, we have that
\bea \label{source}&&\left[F_3\right] \sim \zeta(3)\Omega_2^{3/2} +\zeta(2)\Omega_2^{-1/2}, \non \\ &&\left[F_4\right] \sim \zeta(2) {\rm ln}\Omega_2, \non \\ &&\left[F_5 \right] \sim \zeta(5) \Omega_2^{5/2} +\zeta(4)\Omega_2^{-3/2}, \non \\ &&\left[F_6\right] \sim \zeta(3)^2 \Omega_2^3 +\zeta(2)\zeta(3)\Omega_2 +\zeta(4)\Omega_2^{-1} +\zeta(6)\Omega_2^{-3}\eea
for the BPS couplings. The structure of $\left[ F_4 \right]$ follows from \C{nonan} and we have written down the one loop contribution. Even though the local interaction $D^2\mathcal{R}^4$ vanishes on--shell, it does contribute to the supersymmetry variation and hence the Poisson equation\footnote{These issues have been discussed in~\cite{Green:2008bf,Basu:2008cf,Basu:2013goa}.}. 

We now analyze transcendentality for some non--BPS couplings with \C{source} as source terms in the Poisson equations\footnote{Some of this analysis that follows have been done in~\cite{Basu:2008cf}, and in a different context, in~\cite{Basu:2013oka}.}. For the $D^8\mathcal{R}^4$ term, we have that
\be \left[ F_7 \right] \sim \left[ F_3 \right] \left[ F_4 \right],\ee
which in the string frame, leads to one and two loop non--analytic contributions schematically given by\footnote{From now onwards, we set the scale of the logarithm to unity for brevity as this plays no role in our analysis.} 
\be \a'^4\zeta(2)\Big[ \zeta(3) + e^{2\phi} \zeta(2)\Big] s^4 {\rm ln}(\a's)\mathcal{R}^4\ee
which preserves transcendentality (i.e., each term has weight 2). For the $D^{10} \mathcal{R}^4$ term, we have that
\be \left[ F_8 \right] \sim \left[ F_3 \right] \left[ F_5 \right] + \left[ F_4 \right]^2\ee
which schematically yields
\be \a'^5\Big[ \zeta(3)\zeta(5) e^{-2\phi} + \zeta(2)\zeta(5) + \zeta(3)\zeta(4)e^{2\phi} + \zeta(6) e^{4\phi} + \zeta(4) e^{2\phi}({\rm ln}(\a's))^2\Big] s^5 \mathcal{R}^4\ee
in the string frame yielding contributions upto three loops. Transcendentality is violated at two and three loops by 1 unit (the non--analytic term at two loops preserves transcendentality). Proceeding similarly for the $D^{12} \mathcal{R}^4$ term, we have 
\be \left[ F_9 \right] \sim \left[ F_3 \right] \left[ F_6 \right] + \left[ F_4 \right] \left[ F_5 \right],\ee  
schematically leading to
\bea &&\a'^6\Big[\zeta(3)^3 e^{-2\phi} + \zeta(2)\zeta(3)^2  +\zeta(3)\zeta(4) e^{2\phi} + \zeta(3)\zeta(6) e^{4\phi} + \zeta(6) e^{4\phi} \non \\&& + \zeta(8) e^{6\phi}    +\zeta(2)\zeta(5) ({\rm ln}(\a's)) +\zeta(6) e^{4\phi} ({\rm ln}(\a's))\Big]  s^6 \mathcal{R}^4\eea
which yields contributions upto four loops in the string frame. While the analytic contributions violate transcendentality by 3 units at three and four loops, the non--analytic contribution violates it by 1 unit at three loops. 

For terms at higher orders in the $\alpha'$ expansion, the source terms in the Poisson equation involve non--BPS couplings. For these couplings, we use the results obtained above to analyze transcendentality. Thus recursively we can perform this analysis for a restricted class of source terms at all orders in the $\alpha'$  expansion. We now consider the first two such cases.   

For the  $D^{14} \mathcal{R}^4$ term, we see that 
\be \label{F10}\left[ F_{10} \right] \sim \left[ F_3 \right] \left[ F_7 \right] + \left[ F_4 \right] \left[ F_6 \right] + \left[ F_5 \right]^2\ee  
which schematically leads to
\bea &&\a'^7\Big[ \zeta(5)^2 e^{-2\phi} + \zeta(5)\zeta(4) e^{2\phi} + \zeta(8)e^{6\phi} + \zeta(3)^2\zeta (2)( {\rm ln}(\a' s)) \non \\ &&+ \zeta(3) \zeta(4) e^{2\phi} ({\rm ln}(\a's)) + \zeta(6)e^{4\phi} ({\rm ln}(\a's)) +\zeta(8) e^{6\phi} ({\rm ln}(\a's))  \Big] s^7\mathcal{R}^4\eea
yielding contributions upto four loops in the string frame. Among the analytic terms, transcendentality is violated by 1 and 2 units at two and four loops respectively, while it is violated by 3 units at four loops among the non--analytic terms. Finally, for the  $D^{16} \mathcal{R}^4$ term, from
\be \left[ F_{11} \right] \sim \left[ F_3 \right] \left[ F_8 \right] + \left[ F_4 \right] \left[ F_7 \right] + \left[ F_5 \right] \left[ F_6 \right]\ee  
we get
\bea &&\a'^8\Big[ \zeta(3)^2 \zeta(5) e^{-2\phi}+ \zeta(2)\zeta(3)\zeta(5) + \zeta(3)^2\zeta(4) e^{2\phi}+\zeta(4)\zeta(5)e^{2\phi} \non \\ &&+\zeta(3)\zeta(6)e^{4\phi} + \zeta(5)\zeta(6) e^{4\phi} +\zeta(8) e^{6\phi} + \zeta(10) e^{8\phi}\non \\ &&+\zeta(3)\zeta(4)e^{2\phi} ({\rm ln}(\a's))^2+ \zeta(6) e^{4\phi} ({\rm ln}(\a's))^2\Big] s^8 \mathcal{R}^4\eea
schematically, yielding contributions upto five loops. Transcendentality is violated by 1 unit at two loops, 1 and 3 units at three loops, 1 unit at four loops and 4 units at five loops\footnote{For the last two cases which involve non--BPS couplings as the source terms in the Poisson equation, we note that possible source terms that we have not considered can change the structure of the non--analytic terms as we have mentioned previously, since they can yield higher powers of ${\rm ln}\Omega_2$ with the same coefficient as the ones we have in our list of source terms, thus changing the value of $p_{max}$ for given $n$. It will be interesting to check if this happens.}. Thus we obtain various contributions for multiloop string amplitudes, and the results agree with known results, and further generalize them.  

Hence proceeding recursively, one can generalize this analysis to higher orders in the $\alpha'$ expansion and obtain the transcendentality of various multiloop amplitudes. Let us analyze an interesting subset of such contributions at all orders in the $\alpha'$ expansion. Restricting ourselves to analytic terms only, we consider maximally transcendentality violating amplitudes at higher orders. From \C{source}, note that it is violated by 1 unit by $\left[ F_5 \right]$ and by 3 units by $\left[ F_6 \right]$. Thus we perform this analysis for $\left[ F_{6k} \right]$ for $6k$ mod 6 ($k \geq 2$) keeping only the maximally transcendentality violating contribution that is analytic in the external momenta\footnote{For $k=1$ we reproduce the results obtained above.}.    

For $\left[ F_{6k} \right]$, the maximally transcendentality violating contribution comes from
\be \left[ F_{6k} \right] \sim \left[ F_{6} \right]^k \sim \zeta(6)^k \Omega_2^{-3k}  \ee
on proceeding recursively, leading to the $3k$ loop contribution
\be \a'^{6k-3}\zeta(6)^k e^{-2(1-3k)\phi} s^{6k-3} \mathcal{R}^4  \ee
in the string frame, which violates transcendentality by $3k$ units. 

For $\left[ F_{6k+1} \right]$, we get only non--analytic contributions which we do not consider. For $\left[ F_{6k+2} \right]$, again proceeding recursively the relevant contribution involves
\be \left[ F_{6k+2} \right] \sim \left[ F_{6} \right]^{k-1} \left[ F_3 \right] \left[ F_5 \right]\sim \zeta(6)^{k-1} \zeta(4) \Omega_2^{3(1-k)} \Big[\zeta(3) +\zeta(2) \Omega_2^{-2}\Big],   \ee
leading to the $3k-1$ and $3k$ loop contributions
\be \a'^{6k-1}\zeta(6)^{k-1} \zeta(4) e^{-2(2-3k)\phi} \Big[\zeta(3) +\zeta(2)e^{2\phi}\Big]s^{6k-1} \mathcal{R}^4\ee
which violates transcendentality by $3k-2$ units. Proceeding similarly, we have that
\be \left[ F_{6k+3} \right] \sim \left[ F_{6} \right]^k \left[ F_3 \right] \sim \zeta(6)^k \Omega_2^{3/2-3k} \Big[\zeta(3) +\zeta(2)\Omega_2^{-2}\Big] \ee
yielding the $3k$ and $3k+1$ loop contributions
\be \a'^{6k}\zeta(6)^k e^{-2(1-3k)\phi} \Big[\zeta(3)+\zeta(2)e^{2\phi}\Big]s^{6k}\mathcal{R}^4  \ee
violating transcendentality by $3k$ units. Also
\be \left[ F_{6k+4} \right] \sim \left[ F_{6} \right]^{k-1} \left[ F_5 \right]^2 \sim \zeta(6)^{k-1}\zeta(8) \Omega_2^{-3k}  \ee
which contributes 
\be \a'^{6k+1}\zeta(6)^{k-1} \zeta(8)e^{6k\phi} s^{6k+1}\mathcal{R}^4\ee
at $3k+1$ loops, violating transcendentality by $3k-1$ units. Finally, we have that
\be \left[ F_{6k+5} \right] \sim \left[ F_{6} \right]^{k} \left[ F_5 \right]  \sim \zeta(6)^{k}\zeta(4) \Omega_2^{-3/2-3k}  \ee
which contributes 
\be \a'^{6k+2}\zeta(6)^k \zeta(4) e^{2(1+3k)\phi} s^{6k+2}\mathcal{R}^4\ee
at $3k+2$ loops, and violates transcendentality by $3k+1$ units. Thus we see how constraints imposed by supersymmetry and S--duality constrains transcendentality of various multiloop amplitudes. These results are consistent with known perturbative results. 

\subsection{Transcendentality at one loop and modular graphs}

From the discussions above, we conclude that transcendentality is violated beyond one loop, with specific patterns of violation. While uniform transcendentality for the tree level amplitude follows from \C{tree}, it is not obvious that it is preserved for the one loop amplitude at all orders in the $\alpha'$ expansion. We now analyze this issue for a restricted set of terms for the analytic as well as non--analytic contributions that result from the one loop amplitude, based on our discussion above. We then study connections with the worldsheet analysis.

\subsubsection{The analytic contributions}

First let us consider the analytic terms that result from the one loop amplitude. From the constraints we have analyzed above,  we see that they involve at least the terms
\bea \label{oneloopan}\zeta(2) \Big[ 1 +\zeta(3) \a'^3 s^3 + \zeta(5) \a'^5 s^5 + \zeta(3)^2 \a'^6 s^6 + \zeta(7) \a'^7 s^7 +\zeta(3)\zeta(5) \a'^8 s^8 +\ldots \Big] \mathcal{R}^4\eea 
which have been denoted schematically. We denote these various contributions as $s^k\mathcal{R}^4$ because for sufficiently large values of $k$, there can be more than one spacetime structure of the form $\s_2^p \s_3^q$ with $2p+3q=k$ and our analysis is insensitive to this distinction\footnote{This happens first for $k=6$, with the two distinct possibilities being $\s_2^3$ and $\s_3^2$.}. While the other terms in \C{oneloopan} have been deduced above, we note that the $\zeta(2)\zeta(7) s^7\mathcal{R}^4$ term  follows from \C{F10} on using $[F_7] \sim \zeta(7)\Omega_2^{7/2}$ which is a consequence of the tree level $D^8\mathcal{R}^4$ term that follows from \C{tree}. We did not obtain this from the source terms in the Poisson equation, and this has to be put in separately\footnote{In the eigenvalue equation, it can arise if the eigenvalue is $35/4$ as a solution to the homogeneous equation~\cite{Basu:2008cf}. This term in $[F_7]$ also  leads to the tree level contribution $\zeta(3)\zeta(7) e^{-2\phi} \a'^7 s^7\mathcal{R}^4$ which is consistent wth \C{tree}.}. 

While one can proceed at higher orders in the $\alpha'$ expansion systematically, let us obtain some one loop contributions at all orders, by considering contributions that arise from $[F_{6k}]$ for $6k$ mod 6 ($ k \geq 2$) (for $k=1$ we reproduce \C{oneloopan}). We consider very simple contributions from the source terms that yield one loop contributions. From $[F_{6k}]$, $[F_{6k+2}]$, $[F_{6k+3}]$, $[F_{6k+4}]$ and $[F_{6k+5}]$ we get the one loop contributions\footnote{We use $\left[F_{6k+4}\right]\sim \left[F_6\right]^{k-1} \left[F_3\right]\left[F_7\right]$ with $[F_7] \sim \zeta(7)\Omega_2^{7/2}$, while for the others, the analysis follows in a straightforward way from the structure of the source terms we have used in obtaining maximally transcendentality violating contributions.}
\bea &&\zeta(2)\zeta(3)^{2k-1} \a'^{6k-3}s^{6k-3} \mathcal{R}^4 , \quad \zeta(2)\zeta(5)\zeta(3)^{2(k-1)} \a'^{6k-1}s^{6k-1} \mathcal{R}^4, \quad \zeta(2)\zeta(3)^{2k} \a'^{6k}s^{6k}\mathcal{R}^4, \non \\ &&\zeta(2)\zeta(7) \zeta(3)^{2(k-1)} \a'^{6k+1} s^{6k+1}\mathcal{R}^4, \quad \zeta(2)\zeta(5)\zeta(3)^{2k-1} \a'^{6k+2}s^{6k+2}\mathcal{R}^4\eea  
respectively, on expressing the quantities in the string frame. It is straightforward to get contributions involving zeta functions of higher weights as well. For example, using
\be [F_{12}] \sim [F_3] [F_9]\ee  
and $[F_9] \sim \zeta(9) \Omega_2^{9/2}$ we get a one loop contribution of the form
\be \zeta(2) \zeta(9) \a'^9s^9\mathcal{R}^4\ee
to the amplitude\footnote{Such a tree level contribution to $[F_9]$ arises if the eigenvalue of the Poisson equation  is $63/4$.}. 

\subsubsection{The non--analytic contributions}

We briefly mention the non--analytic contributions as the analysis is similar to what we have done before. Based on the constraints of supersymmetry and S--duality, we see that the first few non--analytic terms beyond $\a's({\rm ln }(\a' s))\mathcal{R}^4$ at one loop are schematically given by
\be \zeta(2) \Big[ \zeta(3) \a'^4s^4 + \zeta(5) \a'^6s^6 + \zeta(3)^2 \a'^7s^7 + \zeta(7) \a'^8 s^8 +\ldots\Big] ({\rm ln} (\a's))\mathcal{R}^4\ee  
among the set of terms we have considered. The basic structure of each of these contributions can be motivated by analyzing the discontinuity across the two particle unitarity cut for the one loop diagram with tree level quartic vertices~\cite{Green:2008uj}. One can generalize this analysis to arbitrary orders in the $\alpha'$ expansion to include certain classes of terms as we have done above for the analytic terms.     

Now all these one loop contributions, both analytic and non--analytic,  preserve transcendentality, suggesting this might be true in general. Also among the set of terms we have considered, ignoring the spacetime structure we see that apart from an overall factor of $\zeta(2)$, these contributions only involve Riemann zeta functions of odd weights and not even weights. This suggests that upto the overall factor or $\zeta(2)$ it is possible that Riemann zeta functions of even weights never appear. Finally, uniform transcendentality of the terms that result from the one loop amplitude also implies that the coefficient of the $\a'^ks^k\mathcal{R}^4$ term has weight one more than the coefficient of the $\a'^ks^k({\rm ln}(\a's)) \mathcal{R}^4$ for every $k$.   

\subsubsection{Transcendentality and the worldsheet analysis}

Even though our spacetime analysis above is only for a class of terms at one loop, we would like to analyze the transcendental structure from the worldsheet perspective as well, in order to obtain connections between the two approaches. Given the limited worldsheet as well as spacetime data, our analysis leads to a specific pattern of contributions to the effective action from the one loop amplitude. 

To begin with, restricting ourselves to local terms in the effective action, they are obtained from  the $\alpha'$ expansion of the one loop amplitude~\cite{Green:1981yb}
\be \label{genus1}\mathcal{A}_{4,an}^{(1)} \sim \pi \mathcal{R}^4\int_{\mathcal{F}_L} \frac{d^2\tau}{\tau_2^2} F(s,t,u;\tau,\bar\tau),\ee
where we have integrated over $\mathcal{F}_L$, the truncated fundamental domain of $SL(2,\mathbb{Z})$ given by~\cite{Green:1999pv}
\be \label{FL}-\frac{1}{2} \leq \tau_1 \leq \frac{1}{2}, \quad \vert \tau \vert \geq 1, \quad \tau_2 \leq L.\ee   
We take $L \rightarrow \infty$ at the end of the calculation and the finite terms that remain give us the local terms in the effective action. The dynamical factor of the amplitude is given by
\be \label{defF}F(s,t,u;\tau,\bar\tau) = \prod_{i=1}^4 \int_{\S}\frac{d^2 z^{(i)}}{\tau_2} e^{\mathcal{D}}\ee
where $z^{(i)}$ ($i=1,\ldots,4$) are the positions of insertions of the vertex operators on the toroidal worldsheet $\S$ of complex structure modulus $\tau$, given by
\be -\frac{1}{2} \leq  {\rm Re} z^{(i)} \leq \frac{1}{2}, \quad 0 \leq {\rm Im} z^{(i)} \leq \tau_2 . \ee
The Koba--Nielsen factor contains $\mathcal{D}$ defined by
\be \a'^{-1}\mathcal{D} \sim  s(G_{12} + G_{34}) + t(G_{14} + G_{23}) + u (G_{13}+ G_{24}),\ee 
where $G_{ij}$ is the scalar Green function between the points $z^{(i)}$ and $z^{(j)}$ and is given by~\cite{Lerche:1987qk,Green:1999pv}
\be G(z;\tau,\bar\tau) = \sum_{(m,n) \neq (0,0)}\frac{\tau_2}{\pi \vert m\tau+n\vert^2}e^{\pi[\bar{z}(m\tau +n)- z(m\bar\tau + n)]/\tau_2}.\ee
Now performing the $\alpha'$ expansion of \C{genus1}, we see that it takes the form
\be \mathcal{A}_{4,an}^{(1)} \sim \pi \mathcal{R}^4\int_{\mathcal{F}_L} \frac{d^2\tau}{\tau_2^2} \Big[ 1+ \sum_{k=2}^\infty \a'^k s^k \mathcal{G}_k(\tau,\bar\tau)\Big]\ee
where we have been schematic about the momentum dependence of the terms in the sum. At order $\alpha'^k$ in the momentum expansion, $\mathcal{G}_k (\tau,\bar\tau)$ is expressible as a sum of a finite number of terms
\be \label{sum1}\mathcal{G}_k (\tau,\bar\tau) = \sum_a \mathcal{G}_k^{(a)} (\tau,\bar\tau), \ee
each of which can be expressed as a product of a finite number of terms
\be \label{prod1}\mathcal{G}_k^{(a)} (\tau,\bar\tau) = \prod_\a \mathcal{G}_k^{(a;\a)} (\tau,\bar\tau).\ee
Every modular graph function $\mathcal{G}_k^{(a;\a)} (\tau,\bar\tau)$ can be depicted by a connected one particle irreducible vacuum graph where the links are given by the scalar Green function, while the vertices are integrated over the torus with measure $d^2z/\tau_2$. Every term $\mathcal{G}_k^{(a)} (\tau,\bar\tau)$ in the sum in \C{sum1} consists of modular graphs with a total of $k$ links and at most four vertices for the four graviton amplitude. For example, $\mathcal{G}_2$ has only one term in the sum given by\footnote{We denote 
\be \int_{ij\ldots} = \int_{\S} \frac{d^2 z^{(i)}}{\tau_2} \int_{\S} \frac{d^2 z^{(j)}}{\tau_2}\ldots\ee
for brevity.}  
\be \mathcal{G}_2  = \int_{12} G_{12}^2, \ee  
while $\mathcal{G}_4$ can be expressed as
\be \mathcal{G}_4 = \sum_{a=1}^3 \mathcal{G}_4^{(a)},\ee
where
\bea \mathcal{G}_4^{(1)} \sim \int_{12} G_{12}^4,\quad \mathcal{G}_4^{(2)} \sim \int_{1234} G_{12} G_{23} G_{34} G_{14}, \quad \mathcal{G}_4^{(3)} \sim (\mathcal{G}_2)^2.\eea

Now while topologically there are several graphs at every order in the $\alpha'$ expansion, there are several algebraic relations between them which vastly reduce the number of them that have to be integrated individually over the moduli space of the complex structure of the torus. This feature, coupled with Poisson equations the modular graphs satisfy on the worldsheet moduli space and their asymptotic expansions around the cusp $\tau_2 \rightarrow \infty$, is helpful in performing these integrals~\cite{DHoker:2015gmr,Basu:2015ayg,DHoker:2015wxz,DHoker:2016mwo,Basu:2016kli,DHoker:2016quv,Kleinschmidt:2017ege,DHoker:2019mib,Basu:2019idd,Gerken:2019cxz} and obtaining the coefficients of the various terms in the effective action. 

Based on known results for the worldsheet analysis for the first few terms in the $\alpha'$ expansion upto the $D^{12}\mathcal{R}^4$ term, a natural pattern emerges for the final expression for the integrand over moduli space at arbitrary orders in the $\alpha'$ expansion\footnote{If there is more than one spacetime structure, this is true for each one of them.}. The integral can be expressed as
\be \label{genint} \pi \a'^k\int_{\mathcal{F}_L} \frac{d^2\tau}{\tau_2^2} \Big[ \Delta \mathcal{F}_k (\tau,\bar\tau)+ c_k + \mathcal{P}_k (\tau,\bar\tau)\Big]s^k\mathcal{R}^4\ee
which contributes to the amplitude at $\O(\alpha'^k)$ in the low momentum expansion for every distinct spacetime structure. 

In \C{genint}, the first term which involves various modular graphs, is a boundary term on moduli space, on using
\be \Delta  = 4\tau_2^2\frac{\p^2}{\p\tau \p \bar\tau}.\ee    
Here $\mathcal{F}_k$ satisfies the Poisson equation
\be \label{ev}\Delta \mathcal{F}_k = \lambda_k \mathcal{F}_k +\ldots,\ee    
where the eigenvalue $\lambda_k$ is non--vanishing. Thus this term contributes
\be \label{c1}\pi \a'^k \int_{-1/2}^{1/2} d\tau_1 \frac{\p \mathcal{F}_k}{\p\tau_2}\Big\vert_{\tau_2 = L \rightarrow \infty}s^k\mathcal{R}^4\ee
to \C{genint}. 

The second term in \C{genint} involves $c_k$ which is a constant, and contributes 
\be \label{c2}2\zeta(2)c_k \a'^k s^k\mathcal{R}^4\ee 
to \C{genint} on integrating over moduli space. Thus \C{c1} and \C{c2} are contributions from the boundary and bulk of moduli space respectively. 

The last term involving $\mathcal{P}_k$ in \C{genint} is given by a sum of terms, each of which is at least bilinear in modular graphs. Though we do not have a detailed understanding of this contribution, available data suggests that on integrating over moduli space and keeping only terms that are constant and those that diverge as ${\rm ln}L $ as $L\rightarrow \infty$, we can choose the graphs that appear in $\mathcal{P}_k$ such that
\be \label{t1}\pi \a'^k\int_{\mathcal{F}_L} \frac{d^2\tau}{\tau_2^2}  \mathcal{P}_k (\tau,\bar\tau)s^k\mathcal{R}^4  \sim \zeta(2) \Big[d_k + e_k ({\rm ln}L)\Big]  \a'^k s^k\mathcal{R}^4,  \ee 
where the constants $d_k$ and $e_k$ have the same transcendentality as\footnote{This condition is important because it does not allow the contribution from $\mathcal{P}_k$  to mix with the contribution from $c_k$ in \C{genint}. In fact, they have different transcendentalities as we shall discuss later.}
\be \label{t2}\frac{1}{\pi}\int_{-1/2}^{1/2} d\tau_1 \frac{\p \mathcal{F}_k}{\p\tau_2}\Big\vert_{\tau_2 = L \rightarrow \infty}.\ee 

Now it is important that in the decomposition in \C{genint}, the eigenvalue in \C{ev} for the Poisson equation for $\mathcal{F}_k$ is non--vanishing. For example, for the $D^6\mathcal{R}^4$ term\footnote{For this interaction $P_3 =0$.}, the integrand involves the modular graph 
\be \int_{12} G_{12}^3\ee 
which satisfies the Poisson equation 
\be \Delta \int_{12} G_{12}^3 = 6 \int_{123} G_{12} G_{23} G_{13}\ee
with vanishing eigenvalue. In this case $\mathcal{F}_3$ can to be expressed in terms of the modular graph $\int_{123} G_{12} G_{23} G_{13}$ using the relation
\be \int_{12} G_{12}^3 = \int_{123} G_{12} G_{23} G_{13} + \zeta(3),\ee
because $\int_{123} G_{12} G_{23} G_{13}$ satisfies the Laplace equation
\be \label{Pe}\Big(\Delta -6 \Big)\int_{123} G_{12} G_{23} G_{13}=0\ee
with non--vanishing eigenvalue. In fact, the shift involving $\zeta(3)$ contributes to $c_3$.

\subsubsection{Connecting the spacetime and worldsheet analysis}

We now analyze the transcendentalities of the various contributions that result from \C{genint} and make contact with the spacetime analysis we have performed earlier. The worldsheet and spacetime analysis is in agreement upto the $D^{12}\mathcal{R}^4$ term upto which the worldsheet analysis has been done. However, we would like to understand how the pattern of the one loop amplitude obtained using \C{genint} compares with the spacetime analysis for interactions at higher orders in the $\alpha'$ interaction. Turning the argument the other way around, the spacetime analysis predicts the transcendentality of certain terms in the worldsheet analysis at all orders in the $\alpha'$ expansion.   

First we consider the consequences of the worldsheet analysis. A modular graph with $l$ links can be expanded around the cusp $\tau_2 \rightarrow \infty$ to yield terms that are power behaved as well as terms that are exponentially suppressed in $\tau_2$. The power behaved terms are of the form~\cite{DHoker:2015gmr,Zerbini:2015rss,DHoker:2015wxz}
\be \label{t3}y^l + \zeta_{sv}(3) y^{l-3} + \zeta_{sv}(5) y^{l-5} +\zeta_{sv}(3)^2 y^{l-6}+ \ldots + \frac{\lambda_{l}}{y^{l-1}},\ee where $y = \pi \tau_2$ , $\lambda_{l}$ has weight $2l-1$ and we have ignored numerical factors of vanishing weight. While each term in \C{t3} has weight $l$, it is conjectured that the coefficient of the $O(y^{l-m})$ term for $m=3$ and $ 5 \leq m \leq 2l-1$ is given by single--valued multiple zeta values of weight $m$, which are defined by a single valued projection of multiple zeta values~\cite{Brown:2013gia}\footnote{Note that 
\be\label{sv}\zeta_{sv}(2n)=0, \quad \zeta_{sv}(2n+1)=2\zeta(2n+1)\ee 
where $n$ is a positive integer greater than 1.}.  

As an aside, we note that the dimension of the space of single--valued multiple zeta values in the ring over the rationals is one upto weight 8, and two at weights 9 and 10~\footnote{For example, at weight 9 the space is spanned by $\zeta_{sv}(9)$ and $\zeta_{sv}(3)^3$, and at weight 10 it is spanned by $\zeta_{sv}(5)^2$ and $\zeta_{sv} (3)\zeta_{sv}(7)$.}. Thus upto weight 10, using \C{sv} we see that they can be expressed in terms of Riemann zeta functions of odd weights. Things are different at weight 11, which will be relevant for our analysis of the five graviton amplitude later. 

Now let us analyze the terms in \C{genint}. To begin with, we see from \C{genus1} and \C{defF} that the integrand is a sum of terms, each of which contains (possibly products of) one particle irreducible vacuum graphs with a total of $k$ links and upto four vertices. Using algebraic relations between the graphs, the integrand can be simplified. Finally, using \C{ev} some of the terms yield the first term in \C{genint}\footnote{Hence we insist that the eigenvalue $\lambda_k$ is non--vanishing.}. Thus $\mathcal{F}_k$ can be expanded as \C{t3} with $l=k$. Expressing $\mathcal{F}_k$ as a sum of terms, this is trivially true for those terms in the sum each of which is a graph with $k$ links. Every term in the sum that is not of this type is a product of $m$ graphs for $m \geq 2$, each with $k_m$ links such that $\sum_m k_m = k$. Each such graph satisfies \C{t3} with $l= k_m$. Thus the product has the asymptotic expansion     
\bea &&\Big(y^{k_1} + \zeta(3) y^{k_1 -3} +\ldots +\frac{\lambda_1}{y^{1-k_1}}\Big)\ldots \Big(y^{k_m} + \zeta(3) y^{k_m -3} +\ldots +\frac{\lambda_m}{y^{1-k_m}}\Big)\non \\ &&=
 y^k +\zeta(3) m y^{k-3}+\ldots +\frac{\lambda_1\ldots \lambda_m}{y^{m-k}}\eea
which is contained in \C{t3} for $l=k$. Thus, suppose the term that is linear in $y$ in $\mathcal{F}_k$ is given by $a_k y$, where $a_k$ has weight $k-1$. Then \C{c1} contributes $\zeta(2) a_k \a'^ks^k\mathcal{R}^4$ to \C{genint}, which has weight one.  
It follows that $c_k$ has transcendentality $k$ and hence \C{c2} has weight two. 
From \C{t2} it follows that $d_k$ and $e_k$ have weight $k-1$, and hence \C{t1} has weight one.  Note that while $a_k$ and $c_k$ are given by single--valued multiple zeta values, we cannot make such a statement in general for $d_k$ and $e_k$ due to our lack of a detailed understanding of \C{t1}. 

Now the non--analytic terms are obtained from the region of the fundamental domain $\mathcal{R}_L$, given by 
\be -\frac{1}{2} \leq \tau_1 \leq \frac{1}{2}, \quad \vert \tau \vert \geq 1, \quad \tau_2 > L,\ee
which is the complement of $\mathcal{F}_L$ in \C{FL}, and thus
\be \label{na}\mathcal{A}_{4,non-an}^{(1)} \sim \pi \mathcal{R}^4\int_{\mathcal{R}_L} \frac{d^2\tau}{\tau_2^2} F(s,t,u;\tau,\bar\tau).\ee  
Since in the total amplitude $\mathcal{A}_{4,an}^{(1)} + \mathcal{A}_{4,non-an}^{(1)}$ the $L$ dependence must cancel, there must be a non-analytic contribution given by\footnote{There are also analytic contributions that arise from \C{na}, and we have assumed they can be included in \C{LN} by appropriately choosing the scale $\m_k$, which can be done at least upto the $s^6\mathcal{R}^4$ term~\cite{DHoker:2019blr}. If this is not the case at higher orders, this will lead to extra analytic terms that have to be analyzed separately. In fact, as mentioned before, unitarity cut construction of the one loop amplitude determines $e_k$ in terms of $\zeta_{sv}(2n+1)$ with $n\geq 1$.}
\be \label{LN} \zeta(2) e_k {\rm ln} (\m_k^{-2}\a's/L) \a'^k s^k\mathcal{R}^4\ee 
which cancels the $L$ dependence of the analytic contribution \C{t1}, and hence the coefficient of this term is completely determined by the analytic term.  

All these terms put together contribute to the $s^k\mathcal{R}^4$ and $({\rm ln}(\a's))s^k\mathcal{R}^4$ terms. This is a sum of weight two terms 
\be \label{w2}\zeta(2) \a'^k \Big[c_k  +  e_k {\rm ln}(\a' \m_k^{-2} s)\Big] s^k\mathcal{R}^4 ,\ee 
and weight one terms
\be \label{w1}\zeta(2) \a'^k \Big[ a_k + d_k\Big] s^k\mathcal{R}^4.\ee   

If we assume that the coefficients that arise in \C{w1} are such that the terms in \C{w1} can be absorbed in the second term in \C{w2} by appropriately choosing the scale $\mu_k$ and then all terms have weight two\footnote{In fact, this is indeed the case upto the $s^6\mathcal{R}^4$ term~\cite{DHoker:2019blr}. } and transcendentality is preserved. However, if this is not the case, this will lead to transcendentality violation at one loop. One can also have violation if the analytic contributions from \C{na} break transcendentality.

It is important to note that the separation between analytic and non--analytic terms is ambiguous because one can always obtain analytic terms from the non--analytic ones by changing the scale of the logarithm. However, there is a natural separation as in \C{w2} because the coefficient of the analytic term has weight one more than that for the non--analytic term. For example, for the $\s_3^2\mathcal{R}^4$ term, $c_6 \sim \zeta(3)^2$, while $e_6 \sim \zeta(5)$.        

Now our spacetime analysis yields contributions that preserve transcendentality, and hence we compare them with the general structure that follows from \C{w2}. 
For the analytic as well as non--analytic contributions, apart from an overall factor of $\zeta(2)$, we obtain only factors of $\zeta_{sv}(2n+1)$ with $n \geq 1$ for the various interactions, which agrees with \C{w2}.    
Also the coefficient of the $\a'^ks^k\mathcal{R}^4$ term has weight $k+2$, which is one more than the coefficient of the $\a'^k s^k({\rm ln}(\a' s))\mathcal{R}^4$ term which has weight $k+1$, in agreement with \C{w2}. 
Thus the spacetime analysis precisely agrees with the structure the follows from the worldsheet analysis for the limited set of interactions we have considered at all orders in the $\alpha'$ expansion.

\section{Analyzing the five graviton amplitude}

We now briefly analyze the issue of transcendentality of multiloop amplitudes for the five graviton amplitude based on constraints of supersymmetry and S--duality. A lot of the analysis is very similar to the four graviton amplitude, hence we shall discuss certain features that are different from the four graviton amplitude. 

At tree level, low momentum expansion of the five graviton amplitude yields contact interactions in the effective action that have coefficients of the Mandelstam variables given by single--valued multiple zeta values~\cite{Stieberger:2009rr,Schlotterer:2012ny,Basu:2013goa,Stieberger:2013wea}. However unlike the four graviton amplitude given by \C{tree} which involves Riemann zeta functions only, higher point amplitudes involve multiple zeta values that cannot be expressed in terms of zeta functions. In the $\alpha'$ expansion of the five graviton amplitude, this first occurs for the $D^{14} \mathcal{R}^5$ term~\cite{Schlotterer:2012ny,Stieberger:2013wea}, which in the string frame, yields terms in the effective action \C{stringframe} of the form
\bea \label{5g}S^{(11)} &\sim& \alpha'^{11} \int d^{10} x \sqrt{-g_\s} e^{-2\phi} \Big[ \zeta_{sv} (3,5,3) (D_\s^{14})_1 + \zeta_{sv}(11) (D_\s^{14})_2\non \\ &&+\zeta_{sv}(3)^2 \zeta_{sv}(5)(D_\s^{14})_3\Big]\mathcal{R}^5.\eea     
The coefficients involve the three elements in the ring over rationals of the basis of single--valued multiple zeta values of weight 11~\cite{Brown:2013gia}. Now using
\be \label{zv} \zeta_{sv}(3,5,3)=2\zeta(3,5,3) - 2\zeta(3) \zeta(5,3) -10 \zeta(3)^2 \zeta(5)\ee
we see that $\zeta_{sv} (3,5,3)$ cannot be expressed in terms of Riemann zeta functions unlike the other coefficients that appear in \C{5g} or at lower orders in the $\alpha'$ expansion\footnote{At weight 11, a basis of multiple zeta values in the ring over rationals has 9 elements which includes the ones on the right hand side of \C{zv}.}. In \C{5g}, $(D_\s^{14})_i \mathcal{R}_\s^5$ ($i=1,2,3$) represents three distinct spacetime structures whose details are not relevant for our purposes\footnote{Note that in~\cite{Schlotterer:2012ny,Stieberger:2013wea}, the basis elements are given by $\zeta_{sv}(5,3,3), \zeta_{sv}(11)$ and $\zeta_{sv}(3)^2 \zeta_{sv}(5)$, while the basis elements we use are the ones in~\cite{Zerbini:2015rss,DHoker:2015wxz}. To go from one to the other, we use 
\be \zeta_{sv}(5,3,3)= 2\zeta(5,3,3) - 5\zeta(3)^2 \zeta(5) + 90 \zeta(2)\zeta(9) +\frac{12}{5} \zeta(2)^2\zeta(7) -\frac{8}{7} \zeta(2)^3 \zeta(5).   \ee
This leads to the relation
\be 2\zeta_{sv}(5,3,3) + \zeta_{sv}(3,5,3) = \frac{299}{2} \zeta_{sv}(11) - \frac{5}{2} \zeta_{sv}(3)^2 \zeta_{sv}(5)\ee
for the desired change in basis, on using~\cite{Blumlein:2009cf}
\be \zeta(3,5,3) + 2\zeta(5,3,3) = \frac{8}{7} \zeta(2)^3 \zeta(5) -\frac{12}{5} \zeta(2)^2\zeta(7)  -90\zeta(2)\zeta(9) + \zeta(3) \zeta(5,3) +\frac{299}{2} \zeta(11).\ee}.      

Proceeding along the lines of the analysis of the four graviton amplitude, we now perform an elementary analysis to obtain some $D^{2k} \mathcal{R}^5$ interactions which contain $\zeta_{sv}(3,5,3)$ beyond tree level\footnote{Low momentum expansion of the five graviton amplitude at one loop has been considered in~\cite{Richards:2008jg,Green:2013bza,Basu:2016mmk}, and at two loops in~\cite{Gomez:2015uha}.}. We include only those source terms in the Poisson equation that result from the $D^{2k'}\mathcal{R}^5$ terms for $k' < k$. While the $\mathcal{R}^5$ term vanishes onshell, the $D^2\mathcal{R}^5$ and $D^4\mathcal{R}^5$ interactions are $1/4$ and $1/8$ BPS respectively.  

For simplicity we consider interactions whose source terms in the Poisson equations have one factor involving the $D^{14}\mathcal{R}^5$ coupling, while the other factor involves the five graviton BPS couplings. 

Considering the $D^2\mathcal{R}^5$ coupling as a factor in the source term, we have that
\be [\widetilde{F}_{16}] \sim [\widetilde{F}_{11}][\widetilde{F}_5],\ee 
leading schematically to the contributions
\be \zeta_{sv}(3,3,5)\a'^{13} \Big[ \zeta(5) e^{-2\phi} +\zeta(4) e^{2\phi}\Big] \tilde{s}^{12}\mathcal{R}^5\ee
at tree level and at two loops in the string frame for the $D^{24}\mathcal{R}^5$ interaction, and thus the two loop term violates transcendentality by 1 unit.           

Similarly considering the $D^4\mathcal{R}^5$ coupling as a factor in the source term, we have that
\be [\widetilde{F}_{17}] \sim [\widetilde{F}_{11}][\widetilde{F}_6],\ee 
leading schematically to the contributions
\be \zeta_{sv}(3,3,5)\a'^{14}\Big[ \zeta(3)^2 e^{-2\phi} +\zeta(2) \zeta(3) + \zeta(4) e^{2\phi} +\zeta(6) e^{4\phi}\Big] \tilde{s}^{13}\mathcal{R}^5\ee
upto three loops in the string frame for the $D^{26}\mathcal{R}^5$ interaction. The three loop term violates transcendentality by 3 units.

\section{Going beyond the type II theory in ten dimensions}

While we have analyzed transcendentality in various amplitudes for the type IIB theory in ten dimensions, it is interesting to consider other cases as well. 

Let us compactify the theory on a circle and consider the type II theory in nine dimensions, and analyze the perturbative contributions to the BPS interactions~\cite{Kiritsis:1997em,Basu:2007ru,Basu:2007ck,Green:2008uj,Basu:2015dqa}. For the $\mathcal{R}^4$ interaction, we get 
\be \Big[ \zeta(3) (re^{-2\phi}) +\zeta(2) \Big(r+\frac{1}{r}\Big)\Big] \mathcal{R}^4\ee
where $r$ is the radius of the circle in the string frame, and as before we have ignored numerical factors with vanishing weight. Assigning vanishing transcendentality to $r$, we see that transcendentality is preserved. For the $D^4\mathcal{R}^4$ interaction we obtain
\be \a'^2\Big[ \zeta(5)(re^{-2\phi}) + \zeta(2)\zeta(3)\Big(r^3+\frac{1}{r^3}\Big) +  \zeta(4) (r e^{-2\phi})^{-1} \Big(r^2+\frac{1}{r^2}\Big)\Big]\s_2\mathcal{R}^4,\ee
and hence transcendentality is violated at one and two loops by 1 unit. Finally the $D^6\mathcal{R}^4$ interaction yields 
\bea &&\a'^3\Big[ \zeta(3)^2 (re^{-2\phi}) + \zeta(2)\zeta(3) \Big(r+\frac{1}{r}\Big) + \zeta(2)\zeta(5) \Big(r^5+\frac{1}{r^5} \Big)\non \\ &&+\zeta(4) (re^{-2\phi})^{-1}\Big(r^2 +\frac{1}{r^2}+\frac{5}{3}\Big)+\zeta(6) (r e^{-2\phi})^{-2}\Big( r^3+\frac{1}{r^3} \Big)\Big] \s_3\mathcal{R}^4.\eea
Thus transcendentality is violated by 2 units at one loop by the term which has $(r^5+r^{-5})$ dependence on the radius of the circle, and by 3 units at three loops. Hence the $1/4$ and $1/8$ BPS interactions violate transcendentality beyond tree level. 

As another example, consider the heterotic theory in ten dimensions. The tree level four graviton amplitude leads to the contact interaction~\cite{Kikuchi:1986rk,Cai:1986sa,Gross:1986mw} 
\be \zeta(3) t_8 t_8 \mathcal{R}^4 +t_8\widetilde{t}_8\mathcal{R}^4\ee
where the two tensor contractions yield different spacetime structures, thus violating transcendentality at tree level.  


\begin{thebibliography}{10}

\bibitem{DHoker:2019blr}
E.~D'Hoker and M.~B. Green, ``{Exploring transcendentality in superstring
  amplitudes},'' {\em JHEP} {\bf 07} (2019) 149,
\href{http://www.arXiv.org/abs/1906.01652}{{\tt 1906.01652}}.

\bibitem{Green:1999pv}
M.~B. Green and P.~Vanhove, ``{The Low-energy expansion of the one loop type II
  superstring amplitude},'' {\em Phys.Rev.} {\bf D61} (2000) 104011,
\href{http://www.arXiv.org/abs/hep-th/9910056}{{\tt hep-th/9910056}}.

\bibitem{Green:2008uj}
M.~B. Green, J.~G. Russo, and P.~Vanhove, ``{Low energy expansion of the
  four-particle genus-one amplitude in type II superstring theory},'' {\em
  JHEP} {\bf 0802} (2008) 020,
\href{http://www.arXiv.org/abs/0801.0322}{{\tt 0801.0322}}.

\bibitem{DHoker:2015gmr}
E.~D'Hoker, M.~B. Green, and P.~Vanhove, ``{On the modular structure of the
  genus-one Type II superstring low energy expansion},'' {\em JHEP} {\bf 08}
  (2015) 041,
\href{http://www.arXiv.org/abs/1502.06698}{{\tt 1502.06698}}.

\bibitem{Green:2006gt}
M.~B. Green, J.~G. Russo, and P.~Vanhove, ``{Non-renormalisation conditions in
  type II string theory and maximal supergravity},'' {\em JHEP} {\bf 02} (2007)
  099,
\href{http://www.arXiv.org/abs/hep-th/0610299}{{\tt hep-th/0610299}}.

\bibitem{Green:2008bf}
M.~B. Green, J.~G. Russo, and P.~Vanhove, ``{Modular properties of two-loop
  maximal supergravity and connections with string theory},'' {\em JHEP} {\bf
  0807} (2008) 126,
\href{http://www.arXiv.org/abs/0807.0389}{{\tt 0807.0389}}.

\bibitem{Basu:2008cf}
A.~Basu and S.~Sethi, ``{Recursion Relations from Space-time Supersymmetry},''
  {\em JHEP} {\bf 09} (2008) 081,
\href{http://www.arXiv.org/abs/0808.1250}{{\tt 0808.1250}}.

\bibitem{Basu:2013goa}
A.~Basu, ``{The structure of the $\mathcal{R}^8$ term in type IIB string
  theory},'' {\em Class.Quant.Grav.} {\bf 30} (2013) 235028,
\href{http://www.arXiv.org/abs/1306.2501}{{\tt 1306.2501}}.

\bibitem{Alday:2018pdi}
L.~F. Alday, A.~Bissi, and E.~Perlmutter, ``{Genus-One String Amplitudes from
  Conformal Field Theory},'' {\em JHEP} {\bf 06} (2019) 010,
\href{http://www.arXiv.org/abs/1809.10670}{{\tt 1809.10670}}.

\bibitem{DHoker:2005jhf}
E.~D'Hoker, M.~Gutperle, and D.~H. Phong, ``{Two-loop superstrings and
  S-duality},'' {\em Nucl. Phys.} {\bf B722} (2005) 81--118,
\href{http://www.arXiv.org/abs/hep-th/0503180}{{\tt hep-th/0503180}}.

\bibitem{DHoker:2014oxd}
E.~D'Hoker, M.~B. Green, B.~Pioline, and R.~Russo, ``{Matching the $D^{6}R^{4}$
  interaction at two-loops},'' {\em JHEP} {\bf 01} (2015) 031,
\href{http://www.arXiv.org/abs/1405.6226}{{\tt 1405.6226}}.

\bibitem{Gomez:2013sla}
H.~Gomez and C.~R. Mafra, ``{The closed-string 3-loop amplitude and
  S-duality},'' {\em JHEP} {\bf 1310} (2013) 217,
\href{http://www.arXiv.org/abs/1308.6567}{{\tt 1308.6567}}.

\bibitem{Beisert:2006ez}
N.~Beisert, B.~Eden, and M.~Staudacher, ``{Transcendentality and Crossing},''
  {\em J. Stat. Mech.} {\bf 0701} (2007) P01021,
\href{http://www.arXiv.org/abs/hep-th/0610251}{{\tt hep-th/0610251}}.

\bibitem{Kotikov:2006ts}
A.~V. Kotikov and L.~N. Lipatov, ``{On the highest transcendentality in N=4
  SUSY},'' {\em Nucl. Phys.} {\bf B769} (2007) 217--255,
\href{http://www.arXiv.org/abs/hep-th/0611204}{{\tt hep-th/0611204}}.

\bibitem{Leurent:2013mr}
S.~Leurent and D.~Volin, ``{Multiple zeta functions and double wrapping in
  planar $N=4$ SYM},'' {\em Nucl. Phys.} {\bf B875} (2013) 757--789,
\href{http://www.arXiv.org/abs/1302.1135}{{\tt 1302.1135}}.

\bibitem{Green:1998by}
M.~B. Green and S.~Sethi, ``{Supersymmetry constraints on type IIB
  supergravity},'' {\em Phys. Rev.} {\bf D59} (1999) 046006,
\href{http://www.arXiv.org/abs/hep-th/9808061}{{\tt hep-th/9808061}}.

\bibitem{Sinha:2002zr}
A.~Sinha, ``{The $\hat{G}^4 \lambda^{16}$ term in IIB supergravity},'' {\em
  JHEP} {\bf 08} (2002) 017,
\href{http://www.arXiv.org/abs/hep-th/0207070}{{\tt hep-th/0207070}}.

\bibitem{Green:1997tv}
M.~B. Green and M.~Gutperle, ``{Effects of D-instantons},'' {\em Nucl. Phys.}
  {\bf B498} (1997) 195--227,
\href{http://www.arXiv.org/abs/hep-th/9701093}{{\tt hep-th/9701093}}.

\bibitem{Green:1999pu}
M.~B. Green, H.-h. Kwon, and P.~Vanhove, ``{Two loops in eleven dimensions},''
  {\em Phys. Rev.} {\bf D61} (2000) 104010,
\href{http://www.arXiv.org/abs/hep-th/9910055}{{\tt hep-th/9910055}}.

\bibitem{Green:2005ba}
M.~B. Green and P.~Vanhove, ``{Duality and higher derivative terms in M
  theory},'' {\em JHEP} {\bf 01} (2006) 093,
\href{http://www.arXiv.org/abs/hep-th/0510027}{{\tt hep-th/0510027}}.

\bibitem{Basu:2013oka}
A.~Basu, ``{Constraining gravitational interactions in the M theory effective
  action},'' {\em Class. Quant. Grav.} {\bf 31} (2014) 165007,
\href{http://www.arXiv.org/abs/1308.2564}{{\tt 1308.2564}}.

\bibitem{Green:1981yb}
M.~B. Green and J.~H. Schwarz, ``{Supersymmetrical String Theories},'' {\em
  Phys.Lett.} {\bf B109} (1982)
444--448.

\bibitem{Lerche:1987qk}
W.~Lerche, B.~E.~W. Nilsson, A.~N. Schellekens, and N.~P. Warner, ``{Anomaly
  Cancelling Terms From the Elliptic Genus},'' {\em Nucl. Phys.} {\bf B299}
  (1988)
91--116.

\bibitem{Basu:2015ayg}
A.~Basu, ``{Poisson equation for the Mercedes diagram in string theory at genus
  one},'' {\em Class. Quant. Grav.} {\bf 33} (2016), no.~5, 055005,
\href{http://www.arXiv.org/abs/1511.07455}{{\tt 1511.07455}}.

\bibitem{DHoker:2015wxz}
E.~D'Hoker, M.~B. Green, O.~Gurdogan, and P.~Vanhove, ``{Modular Graph
  Functions},'' {\em Commun. Num. Theor. Phys.} {\bf 11} (2017) 165--218,
\href{http://www.arXiv.org/abs/1512.06779}{{\tt 1512.06779}}.

\bibitem{DHoker:2016mwo}
E.~D'Hoker and M.~B. Green, ``{Identities between Modular Graph Forms},'' {\em
  J. Number Theor.} {\bf 189} (2018) 25--88,
\href{http://www.arXiv.org/abs/1603.00839}{{\tt 1603.00839}}.

\bibitem{Basu:2016kli}
A.~Basu, ``{Proving relations between modular graph functions},'' {\em Class.
  Quant. Grav.} {\bf 33} (2016), no.~23, 235011,
\href{http://www.arXiv.org/abs/1606.07084}{{\tt 1606.07084}}.

\bibitem{DHoker:2016quv}
E.~D'Hoker and J.~Kaidi, ``{Hierarchy of Modular Graph Identities},'' {\em
  JHEP} {\bf 11} (2016) 051,
\href{http://www.arXiv.org/abs/1608.04393}{{\tt 1608.04393}}.

\bibitem{Kleinschmidt:2017ege}
A.~Kleinschmidt and V.~Verschinin, ``{Tetrahedral modular graph functions},''
  {\em JHEP} {\bf 09} (2017) 155,
\href{http://www.arXiv.org/abs/1706.01889}{{\tt 1706.01889}}.

\bibitem{DHoker:2019mib}
E.~D'Hoker, ``{Integral of two-loop modular graph functions},'' {\em JHEP} {\bf
  06} (2019) 092,
\href{http://www.arXiv.org/abs/1905.06217}{{\tt 1905.06217}}.

\bibitem{Basu:2019idd}
A.~Basu, ``{Eigenvalue equation for the modular graph $C_{a,b,c,d}$},'' {\em
  JHEP} {\bf 07} (2019) 126,
\href{http://www.arXiv.org/abs/1906.02674}{{\tt 1906.02674}}.

\bibitem{Gerken:2019cxz}
J.~E. Gerken, A.~Kleinschmidt, and O.~Schlotterer, ``{All-order differential
  equations for one-loop closed-string integrals and modular graph forms},''
\href{http://www.arXiv.org/abs/1911.03476}{{\tt 1911.03476}}.

\bibitem{Zerbini:2015rss}
F.~Zerbini, ``{Single-valued multiple zeta values in genus 1 superstring
  amplitudes},'' {\em Commun. Num. Theor. Phys.} {\bf 10} (2016) 703--737,
\href{http://www.arXiv.org/abs/1512.05689}{{\tt 1512.05689}}.

\bibitem{Brown:2013gia}
F.~Brown, ``{Single-valued Motivic Periods and Multiple Zeta Values},'' {\em
  SIGMA} {\bf 2} (2014) e25,
\href{http://www.arXiv.org/abs/1309.5309}{{\tt 1309.5309}}.

\bibitem{Stieberger:2009rr}
S.~Stieberger, ``{Constraints on Tree-Level Higher Order Gravitational
  Couplings in Superstring Theory},'' {\em Phys.Rev.Lett.} {\bf 106} (2011)
  111601,
\href{http://www.arXiv.org/abs/0910.0180}{{\tt 0910.0180}}.

\bibitem{Schlotterer:2012ny}
O.~Schlotterer and S.~Stieberger, ``{Motivic Multiple Zeta Values and
  Superstring Amplitudes},'' {\em J. Phys.} {\bf A46} (2013) 475401,
\href{http://www.arXiv.org/abs/1205.1516}{{\tt 1205.1516}}.

\bibitem{Stieberger:2013wea}
S.~Stieberger, ``{Closed superstring amplitudes, single-valued multiple zeta
  values and the Deligne associator},'' {\em J. Phys.} {\bf A47} (2014) 155401,
\href{http://www.arXiv.org/abs/1310.3259}{{\tt 1310.3259}}.

\bibitem{Blumlein:2009cf}
J.~Blumlein, D.~J. Broadhurst, and J.~A.~M. Vermaseren, ``{The Multiple Zeta
  Value Data Mine},'' {\em Comput. Phys. Commun.} {\bf 181} (2010) 582--625,
\href{http://www.arXiv.org/abs/0907.2557}{{\tt 0907.2557}}.

\bibitem{Richards:2008jg}
D.~M. Richards, ``{The One-Loop Five-Graviton Amplitude and the Effective
  Action},'' {\em JHEP} {\bf 0810} (2008) 042,
\href{http://www.arXiv.org/abs/0807.2421}{{\tt 0807.2421}}.

\bibitem{Green:2013bza}
M.~B. Green, C.~R. Mafra, and O.~Schlotterer, ``{Multiparticle one-loop
  amplitudes and S-duality in closed superstring theory},'' {\em JHEP} {\bf 10}
  (2013) 188,
\href{http://www.arXiv.org/abs/1307.3534}{{\tt 1307.3534}}.

\bibitem{Basu:2016mmk}
A.~Basu, ``{Simplifying the one loop five graviton amplitude in type IIB string
  theory},'' {\em Int. J. Mod. Phys.} {\bf A32} (2017), no.~14, 1750074,
\href{http://www.arXiv.org/abs/1608.02056}{{\tt 1608.02056}}.

\bibitem{Gomez:2015uha}
H.~Gomez, C.~R. Mafra, and O.~Schlotterer, ``{Two-loop superstring five-point
  amplitude and $S$-duality},'' {\em Phys. Rev.} {\bf D93} (2016), no.~4,
  045030,
\href{http://www.arXiv.org/abs/1504.02759}{{\tt 1504.02759}}.

\bibitem{Kiritsis:1997em}
E.~Kiritsis and B.~Pioline, ``{On $R^4$ threshold corrections in type IIB
  string theory and (p,q) string instantons},'' {\em Nucl. Phys.} {\bf B508}
  (1997) 509--534,
\href{http://www.arXiv.org/abs/hep-th/9707018}{{\tt hep-th/9707018}}.

\bibitem{Basu:2007ru}
A.~Basu, ``{The $D^4 R^4$ term in type IIB string theory on $T^2$ and U-
  duality},'' {\em Phys. Rev.} {\bf D77} (2008) 106003,
\href{http://www.arXiv.org/abs/0708.2950}{{\tt 0708.2950}}.

\bibitem{Basu:2007ck}
A.~Basu, ``{The $D^6 R^4$ term in type IIB string theory on $T^2$ and U-
  duality},'' {\em Phys. Rev.} {\bf D77} (2008) 106004,
\href{http://www.arXiv.org/abs/0712.1252}{{\tt 0712.1252}}.

\bibitem{Basu:2015dqa}
A.~Basu, ``{Perturbative type II amplitudes for BPS interactions},'' {\em
  Class. Quant. Grav.} {\bf 33} (2016), no.~4, 045002,
\href{http://www.arXiv.org/abs/1510.01667}{{\tt 1510.01667}}.

\bibitem{Kikuchi:1986rk}
Y.~Kikuchi, C.~Marzban, and Y.~J. Ng, ``{Heterotic String Modifications of
  Einstein's and {Yang-Mills}' Actions},'' {\em Phys. Lett.} {\bf B176} (1986)
57--60.

\bibitem{Cai:1986sa}
Y.~Cai and C.~A. Nunez, ``{Heterotic String Covariant Amplitudes and Low-energy
  Effective Action},'' {\em Nucl. Phys.} {\bf B287} (1987)
279.

\bibitem{Gross:1986mw}
D.~J. Gross and J.~H. Sloan, ``{The Quartic Effective Action for the Heterotic
  String},'' {\em Nucl. Phys.} {\bf B291} (1987)
41--89.

\end{thebibliography}

\providecommand{\href}[2]{#2}\begingroup\raggedright\endgroup

\end{document}